# Electron-phonon interaction, magnetic phase transition, charge density waves and resistive switching in VS$_2$ and VSe$_2$ revealed by Yanson point contact spectroscopy

D. L. Bashlakov[1], O. E. Kvitnitskaya[1,2], S. Aswartham[2], G. Shipunov[2], L. Harnagea[3], D. V. Efremov[2], B. Büchner[3,4], Yu. G. Naidyuk[1]

[1]B. Verkin Institute for Low Temperature Physics and Engineering, NAS of Ukraine, 61103 Kharkiv, Ukraine

[2]Leibniz Institute for Solid State and Materials Research, IFW Dresden, 01069 Dresden, Germany

[3]I-HUB Quantum Technology Foundation, Indian Institute of Science Education and Research (IISER), Pune 411008, India

[4] Institute of Solid State and Materials Physics and Würzburg-Dresden Cluster of Excellence ct.qmat, Technische Universität Dresden, 01062 Dresden, Germany

**Abstract**

VS$_2$ and VSe$_2$ have attracted particular attention among the transition metals dichalcogenides because of their promising physical properties concerning magnetic ordering, charge density wave (CDW), emergent superconductivity, etc., which are very sensitive to stoichiometry and dimensionality reduction. Yanson point contact (PC) spectroscopic study reveals metallic and nonmetallic states in VS$_2$ PCs, as well as a magnetic phase transition was detected below 25 K. Analysis of PC spectra of VS$_2$ testifies the realization of the thermal regime in PCs. At the same time, rare PC spectra, where the magnetic phase transition was not visible, shows a broad maximum of around 20 mV, likely connected with electron-phonon interaction (EPI). On the other hand, PC spectra of VSe$_2$ demonstrate metallic behavior, which allowed us to detect features associated with EPI and CDW transition. The Kondo effect appeared for both compounds, apparently due to interlayer vanadium ions. Besides, the resistive switching was observed in PCs on VSe$_2$ between a low resistive, mainly metallic-type state, and a high resistive nonmetallic-type state by applying bias voltage (about 0.4V). In contrast, reverse switching occurs by applying a voltage of opposite polarity (about 0.4V). The reason may be the alteration of stoichiometry in the PC core due to the displacement of V ions to interlayer under a high electric field. The observed resistive switching characterize VSe$_2$ as a potential material, e.g., for non-volatile resistive RAM, neuromorphic engineering, and for other nanoelectronic applications. At the same time, VSe$_2$ attracts attention as a rare layered van der Waals compound with magnetic transition.

**Introduction**

Layered transition metal dichalcogenides (TMDs) are class of materials characterized by their unique structure, composed of stacked two-dimensional layers held together by relatively weak van der Waals forces, while intra-layer strength is protected by strong covalency. This layered arrangement gives rise to anisotropic properties, where the physical characteristics can significantly vary along different crystallographic directions. Depending on the arrangement of the atoms, the structures of TMDs can be categorized as trigonal prismatic (hexagonal, H), octahedral (tetragonal, T), and their distorted phase T`. Depending on their chemical compositions and structural configurations, TMDs can be classified as metallic, semimetallic, and semiconducting with varied band structures, as well, they can demonstrate charge density waves (CDW), diversity of phase transition, and superconductivity [1].

In recent years, significant progress has been made in understanding and exploiting the unique electronic properties of these materials [2]. Advanced experimental techniques, such as angle-resolved photoemission spectroscopy, scanning tunneling microscopy/spectroscopy and transport measurements, combined with theoretical modeling and simulations, have shed light on the underlying mechanisms governing the electronic behavior of layered chalcogenides.

TMDs $VS_2$ and $VSe_2$, composed of vanadium and chalcogen atoms, possess intriguing electronic properties that have attracted the attention of scientists in recent years. Interesting, that vanadium dichalcogenides can be in semiconducting as well as in metallic phases. However, the attractiveness of these two materials lies in the fact that they demonstrate inherent (ferro)magnetic properties [3], which can significantly expand their potential use as highly efficient functional nanomaterials that actually revolutionize the development of spintronics.

The electronic properties of layered chalcogenides, including $VS_2$ and $VSe_2$, can be further tailored through various methods, such as chemical doping, alloying, or heterostructure engineering. These approaches allow researchers to manipulate the band structure, modify the charge transport characteristics, and explore new device design and functionality possibilities.

Znang *et al*. [4] concluded that bulk and monolayer of $VS_2$ in the H- and T-phase are stable, while they prefer to exhibit the hexagonal H-structure instead of the trigonal T-structure below room temperature. Gauzzi *et al.* [5] found that $1T$ - $VS_2$ is nonmetallic and displays no long-range structural modulations, such as CDW, contrary to marked metallic behavior of the $1T$ - $VS_2$ sample below 260 K with CDW transition around 300 K, reported by Mulazzi *et al*. [6].

It is well known that the transport properties of TMDs strongly depend on their stoichiometry. However, as many studies have reported, it is challenging to grow stoichiometric $VS_2$ because of self intercalation. Thus, according to Ref. [7], self-intercalated by V atoms in the van der Waals gap 1T-$V_{1+x}S_2$ single crystals with controlled concentration, x = 0.09−0.17 are metallic up to 400 K, with no CDW order. These authors also claim that the stoichiometric (x = 0) phase of $VS_2$ remains little studied because of its thermodynamic metastability, as explained by *ab initio* calculations, so, the stable phase is, e.g., x = 0.25 (or $V_5S_8$) with the Néel temperature $T_N$=32K [8]. Variation of $VS_2$ properties was observed also for nanosheets with sub-10 nm thickness [9], where 50 nm film shows metallic type of resistivity $\rho(T)$ similar to a mentioned phase with x = 0.25 (or $V_5S_8$), but $\rho(T)$ of thinner 6.7 nm film demonstrates nonmetallic behavior resembling data shown in

[5]. Also, recent experiments by Niu *et al.* [10] point to important role of Kondo physics in the intercalated $V_5S_8$ suggesting that itinerant electrons are correlated with enhanced by factor about 1.6 bare electron mass. Thus, the $V_5S_8$ can be attributed to the category of correlated electronic systems, where a Kondo lattice may even emerge.

Let`s turn to $VSe_2$. The bulk crystal $VSe_2$ of 1T-phase has a metallic behavior, while structural phase transition of $VSe_2$ multilayers can take place from 1T to semiconducting 2H- phase through annealing at 650 K, which is more thermodynamically 2D favorable than the 1T-phase according to Li *et al.* [11]. An interesting feature in the resistivity $\rho(T)$ of 1T-$VSe_2$ is its quadratic temperature dependence observed below 50 K along with a bend in $\rho(T)$ around 100 K [12], which is associated with CDW transition [13]. In addition, signatures of the Kondo effect were found in $VSe_2$ [14] due to presence of magnetic interactions between the paramagnetic interlayer V ions and a Kondo screening of these V moments.

There are also conflicting reports as to superconductivity in $VSe_2$ under pressure. Sahoo *et al.* [15] announced that superconducting transition emerges at about ∼4 K at pressure ∼15 GPa, while Song *et al*. [16] reported that the CDW transition shifts gradually to higher temperature with increasing pressure, but either quasi-hydrostatic or non-hydrostatic compression, no sign of superconductivity is observed from 2 to 300 K up to 42 GPa.

Accounting for plethora of ground states and size dependent characteristics of $VS_2$ and $VSe_2$, along with their emerging magnetic properties, it was challenging to apply Yanson point-contact (PC) spectroscopy [17] to their investigation. Recently, we studied by this method series of other TMDs, such as $MoTe_2$, $WTe_2$, Ta*Me*$Te_4$ (*Me*= Ru, Rh, Ir) [18, 19, 20], $TiSe_2$, TiSeS, $Cu_xTiSe_2$ [21] and $TiTe_2$ [22], where such phenomena as electron-phonon interaction (EPI), emergent superconductivity, phase transition, resistive switching etc. in restricted PC geometry under high current density and strong electric field were observed and investigated.

**Experimental details**

*Samples*. The single-crystals of $VS_2$ were grown by Chemical Vapour Transport technique. Shiny plate-like crystals up to few millimeter in size were obtained with layered morphology [10]. The as grown crystals show the off-stoichiometry $V_{1.1}S_2$ with excess V atoms are intercalated in-between the layers.

The single crystal growth of $VSe_2$ and the electrical characterization of crystals were described in detail in the Supplement of [15]). We used $VSe_2$ crystals shaped as thin plates with lateral dimensions ranging from only few mm to as large as 15x10x0.1 mm with RRR ($R_{300K}/R_{10K}$) about 5. They demonstrate resistance inflection near CDW transition at 110K and strong anisotropic resistance. Observed resistance upturn below 15 K in $R_{ab}$ was related to Kondo scattering from interlayer localized V magnetic moments. Above the upturn, between 15 and 36 K, the resistance follows $R_{ab}=R_0+AT^n$ dependence with n=2.93.

*Point contact measurements.* PCs were prepared by touching of a thin Ag or Cu wire to the flat surface of a flake, cleaved at room temperature, or by contacting its edge/side with this wire. Also, so-called "soft" PCs were made by dripping a small drop of silver paint onto the cleaned/cleaved sample surface/edge. Thus, we carried out resistive measurements on the hetero-contacts between a simple metal (Ag, Cu or silver paint) and the investigated samples. We measured the current–voltage *I(V)* characteristics of PCs and their first derivatives *dV/dI(V)* and second derivatives *d²V/dI²(V)*. The derivatives of *I(V)* curves were recorded by scanning the *dc* current *I* through PC, on which a small *ac* current *i* was superimposed, and detecting the first and second harmonic of *ac* signal using a standard lock-in technique. To look for the switching effect, we swept voltage back and forth across the point contact with increasing its amplitude until resistance switching was observed. The measurements were carried out in the temperature range from liquid helium up to room temperature and at a magnetic field up to 15T.

### Experimental results and discussion.

*dV/dI(V)* PC spectra of $VS_2$ can be divided into two groups. The first one demonstrates nonmetallic *dV/dI(V)* behavior at low temperatures with a sharp zero-bias peak (see Fig. 1(a)). The peak is suppressed with increasing temperature and disappears around 25 K. At the same time, a magnetic field up to 15 T has no significant impact on this peak (see insets in Fig. 1). As it can be seen from Fig. 1(b), the peak has log-behavior between 0.6-6 mV.

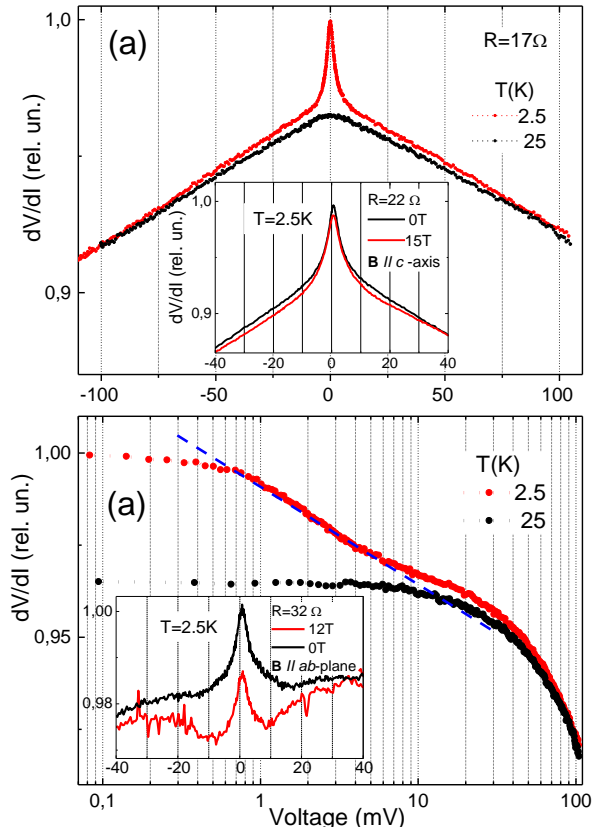

Fig.1. (a) Behavior of nonmetallic-type *dV/dI(V)* of $VS_2$ – Ag PC at different temperatures. (b) *dV/dI(V)* from the upper panel in the log-scale. Both insets show behavior of *dV/dI(V)* in a magnetic field.

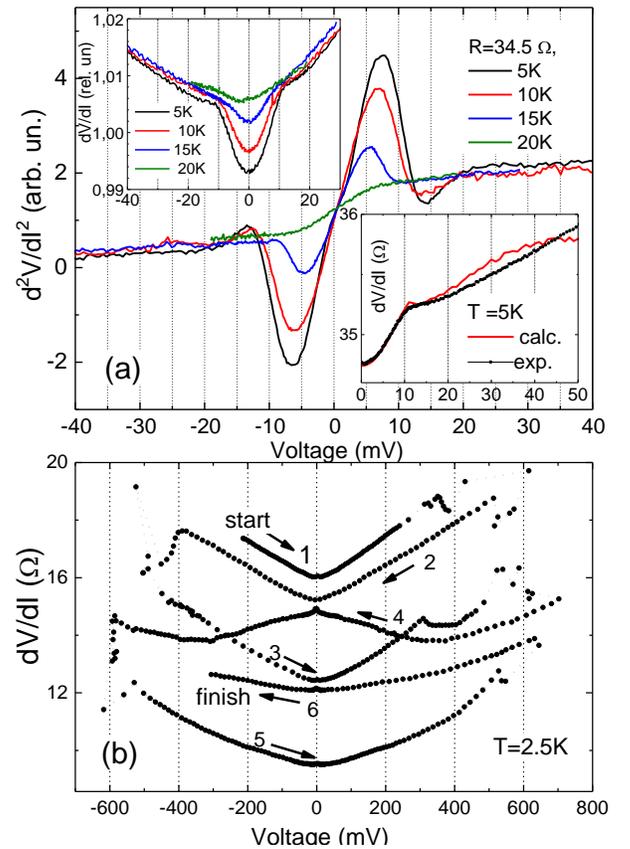

Fig. 2. (a) Behavior of *d²V/dI²(V)* and *dV/dI(V)* (inset) of metallic-type PC of $VS_2$ – Ag at different temperatures. The bottom inset shows *dV/dI(V)* at 5K compared to the calculation (see text for details). (b) Evolution of *dV/dI(V)* of $VS_2$ – Ag PC at larger bias by repeated scanning of voltage forth and back from curve 1 to 6.

The second group of *dV/dI(V)* spectra shows metallic behavior with parabolic-like minimum on which an addition minimum develops below 25 K (see inset in Fig. 2(a)). Such transformation of *dV/dI(V)* with temperature is more clearly seen in the second derivatives $d^2V/dI^2(V)$ (see in Fig. 2(a) and 3). Figure 3 also demonstrates in more details for another PC the evolution of the main peak in $d^2V/dI^2(V)$ with increasing of temperature and magnetic fields. In both cases, this structure is gradually suppressed by temperature rise and vanishes above 20 K or in a magnetic field above 10 T. Contrary, Fig. 4 presents a rare $d^2V/dI^2(V)$ spectrum with a broad maximum around 20 mV, which survives with temperature rise above 20 K, while zero-bias N-shaped anomaly is suppressed. Interestingly is, this maximum is in the region of characteristic phonon energies of $VS_2$ [4, 5], and it is connected with EPI in $VS_2$.

We tried to search the switching effect in PCs on $VS_2$, similarly as we did in Refs. [20, 21, 22], sweeping voltage to the higher values. As shown in Fig. 2(b), we have observed PC resistance jumps, but not repetitive states.

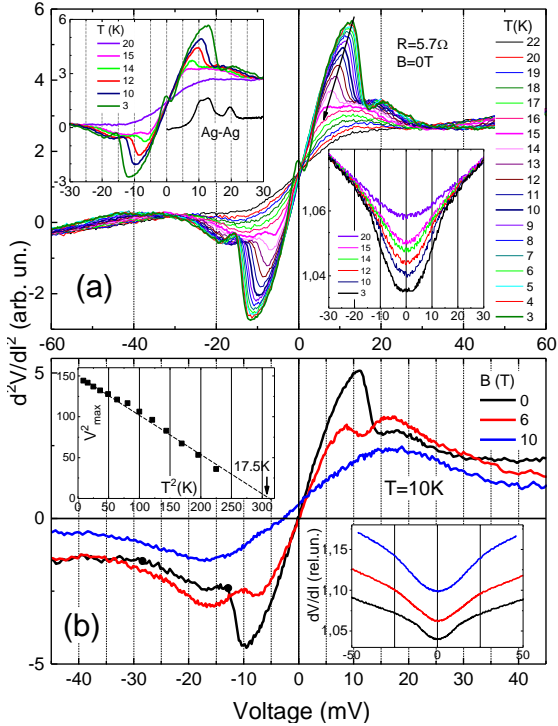

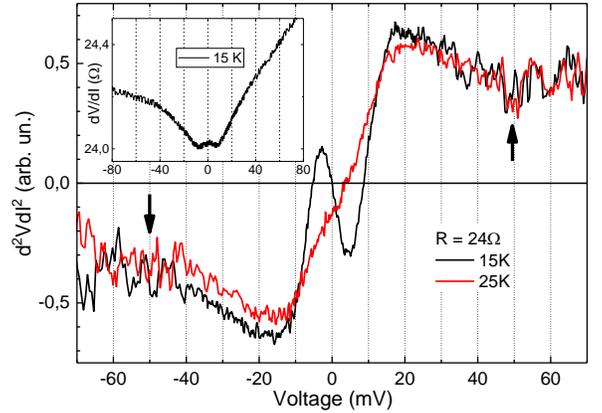

Fig. 3. (a) Behavior of another metallic-type *dV/dI(V)* and $d^2V/dI^2(V)$ of $VS_2$– Ag PC at different temperatures. Insets show *dV/dI(V)* and $d^2V/dI^2(V)$ at selected temperatures. (b) The behavior of *dV/dI(V)* and $d^2V/dI^2(V)$ of $VS_2$– Ag PC at 10 K from the upper panel in a magnetic field. The upper inset shows the position of the main maximum in $d^2V/dI^2(V)$ from the upper panel versus temperature in $V^2$ vs $T^2$ coordinates. The bottom inset shows *dV/dI(V)* for PC from the main panel.

Fig. 4. $d^2V/dI^2(V)$ spectra of PC $VS_2$– Ag PC at temperatures 15 and 25 K. Inset shows *dV/dI(V)* at 15 K. Arrows mark maximal phonon energy in $VS_2$ according to [4, 5].

Concerning spectra in Fig. 1 with nonmetallic behavior of *dV/dI(V)* and sharp zero bias peak. As mentioned by Gauzzi et al. [5], 1T- $VS_2$ demonstrates nonmetallic behavior of resistivity and displays no long-range structural modulations, such as CDW. Subsequently, Moutaabbid et al. [7] reported that the metallic type phase of $VS_2$ appears with excess V. In addition, Ji et al. [9] demonstrated measurements on crystalline $VS_2$ nanosheets with "excellent" electrical conductivities, where their resistance behavior changes from metallic to nonmetallic by decreasing their thicknesses from 50 nm to 6.7 nm. Thus, all these show a strong dependence of $VS_2$ properties on sample preparation method (quality and stoichiometry), which, in our case, results in variation of *dV/dI(V)* behavior in different PCs. Thus, the nonmetallic type of our *dV/dI(V)* curves may be due to a more stoichiometric condition of $VS_2$ in the PC core. And vice versa, the metallic type of *dV/dI(V)* is characteristic for PCs with excess of interlayer V atoms.

Let us consider the origin of minimum in metallic type *dV/dI(V)*, which appears below 25 K and deepens with decreasing temperature (see Figs. 2 & 3). Yanson PC spectroscopy considers three regimes of electron transport in PCs: ballistic, diffusive, and thermal [17, 23]. The first two take place in PCs when the inelastic mean free path of electrons is larger than the contact size. Only in this case spectroscopy is possible, while electrons gain excess energy exactly equal to eV (e is the electron charge) by applying voltage V to PC and can create quasiparticles with a certain energy by scattering. On the contrary, in the thermal regime, when the inelastic mean free path of electrons is smaller than the contact size, electrons lose their excess energy within PC, and Joule heating occurs. In the case of the fulfillment of Wiedemann-Franz law, the temperature in the PC ($T_{PC}$) in this regime increases with a bias voltage as [24]:

$$T^2_{PC} = T^2_{bath} + V^2/4L, \qquad (1)$$

where $T_{bath}$ is the temperature of environment, $L = \pi^2 k_B^2/3e^2$ is the Lorenz number. At high voltage $eV \gg k_B T_{bath}$ or at low temperature $T_{bath}$, $T_{PC}$ rises linearly with a voltage with a rate 3.2 K/mV for standard Lorenz number $L_0 = 2.45 \cdot 10^{-8}$ $V^2/K^2$.

Here we must note that the inelastic mean free path of electrons in PC depends on voltage. At low voltages, electrons do not get enough energy to create quasiparticles and scatter, but with increasing voltage, the probability of scattering increases and the inelastic mean free path begins to decrease. That is the same PC can demonstrate spectral characteristics at low voltages and transition to a thermal regime with voltage increase as was shown in the example of ordinary ferromagnetic metals [24].

In general, in the thermal regime, *dV/dI(V)* mimics resistivity *ρ(T)* behavior considering the relation between *V* and *T* according to Eq. (1). To be more accurate, *dV/dI(V)* can be calculated from the equation for *I(V)* curve, which was obtained in the thermal regime [24]:

$$I(V) = Vd \int_0^1 \frac{dx}{\rho(T_{PC}(1-x^2)^{1/2})}, \qquad (2)$$

where *d* is the PC diameter and *ρ(T)* is the resistivity. As follows from the literature data, *ρ(T)* in $VS_2$ strongly depends on the quality of samples, e.g., stoichiometry [7], and varies for thin films [9]. From these two references, we found that *dV/dI(V)* (see inset in Fig. 2(a)) is similar to *ρ(T)* of $V_5S_8$

(or $V_{1.25}S_2$) [8] or to $\rho(T)$ of 50 nm thick film [9], where both of curves show a kink in $\rho(T)$ attributed to magnetic transition.

So, to calculate $dV/dI(V)$ from Eq. (2), we used the resistivity $\rho(T)$ of 50 nm thin film from [9]. To fit the position of the kink in $dV/dI(V)$, we used Lorenz number $L=1.2L_0$. To describe the minimum depth, we alter residual resistivity to $\rho_0 = 5.7$ mΩ·cm, and to get correct zero-bias resistance of PC, we used diameter d=1.7 μm. As a result, the theoretical curve in the inset of Fig. 2(b) describes quite reasonably the experimental $dV/dI(V)$. Thus, we confirm the realization of the thermal regime by calculations. Also, the position of distinct peculiarities like kink (maximum) in $dV/dI(V)$ ($d^2V/dI^2(V)$) shifts with temperature linearly versus temperature in $V^2$ vs $T^2$ coordinates according to Eq. (1), as it is seen in inset of Fig. 3(b).

The magnetic origin of the minimum in $dV/dI(V)$ or the maximum in $d^2V/dI^2(V)$ confirms their suppression by magnetic field in Fig. 3(b). Similar suppression of peak in $d^2V/dI^2(V)$ by temperature and in magnetic field was observed in PCs for another material with antiferromagnetic transition, namely, for nickel borocarbide $HoNi_2B_2C$ with $T_N$= 5 K (see Fig. 5 in [25]). Suppression of broad minimum in $dV/dI(V)$ with temperature rise above 100 K was also reported for PCs on iron-based superconductor $Ba_{1-x}Na_xFe_2As_2$ with antiferromagnetic state below 120 K (see Fig. 1 in [26]).

Returning to the spectra in Fig. 4, let us notice that $d^2V/dI^2(V)$ represents the EPI function in Yanson PC spectroscopy [17]. Therefore, the broad maximum around 20 mV followed by a featureless background above maximal phonon energy (marked by arrows) in $VS_2$ [4, 5] reflects EPI in this material.

$dV/dI(V)$ PC spectra of $VSe_2$ are shown in Figs. 5 and 6. $dV/dI(V)$ demonstrates parabolic-like dependence at low voltages with zero bias maximum (inset of Fig. 5(a) ), which intensity varies for different PCs (compare $dV/dI(V)$ in the inset of Fig. 5(a) with Fig. 6(a)). This maximum is very likely connected with the Kondo effect [17, 28], due to the presence of interlayer V ions, which can provide the localized magnetic moments [14, 15, 29]. Zero-bias maximum on $dV/dI(V)$ results in a pronounced N-shaped feature in $d^2V/dI^2(V)$ (see Fig. 5(b)). The latter also contains a blurred structure, which transforms into a broad maximum at around 20 mV after the subtraction of the linear background shown by the dashed line in Fig. 5(b). It is logical to use such type of background for $d^2V/dI^2(V)$, considering that $dV/dI(V)$ can be fitted well by parabola (see inset in Fig. 5(a)). The blurred structure vanishes above 40 mV, which corresponds to the maximal phonon energy in $VSe_2$, according to [27]. The position of the broad maximum around 20 mV and a less pronounced shoulder/kink around 10 mV correspond well to the main maximum and to several sharp maxima around 10 meV in the EPI function presented in [27]. Therefore, it is natural to suppose that the observed structure in $d^2V/dI^2(V)$ reflects an EPI in $VSe_2$. It is interesting that the main peak of EPI function in clean vanadium is also at 20 meV, and the maximal phonon energy is about 35 meV [30].

Yanson PC spectroscopy was used by Kamarchuk *et al.* [31] to study EPI in $1T-VSe_2$. Their $d^2V/dI^2(V)$ demonstrates a huge peak around 6 mV followed by non-conventional as for EPI PC spectra puzzle structure up to 90 meV, which is more than twice as much as the maximal phonon energy, according to [27]. Therefore, we have doubts about their interpretation of these data as related to EPI in $VSe_2$. The authors of Ref. [31] also show $d^2V/dI^2(V)$ similar to that in Fig. 5(b) with close to zero-bias minimum (they demonstrate $d^2V/dI^2(V)$ only for one polarity), however, they interpret such a curve as "inverse" EPI spectrum. As mentioned above, this feature corresponds to zero-bias maximum in $dV/dI(V)$ and is likely connected with a Kondo effect in VSe$_2$ observed in [14, 15, 29].

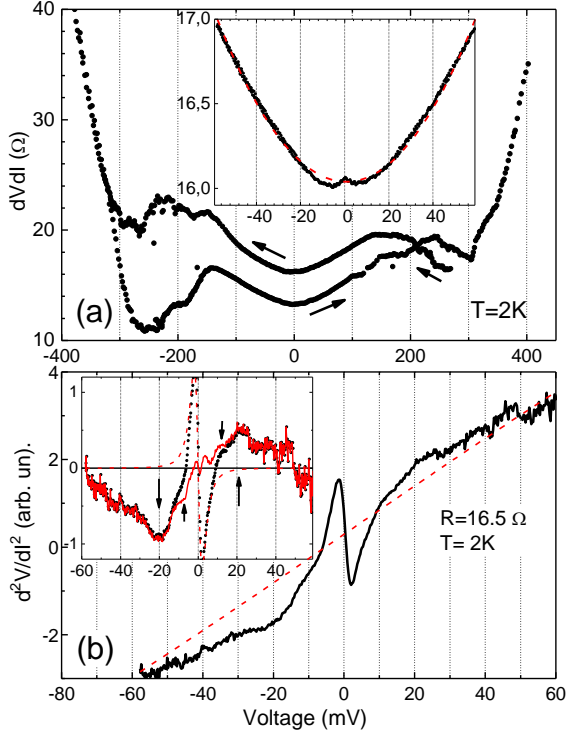

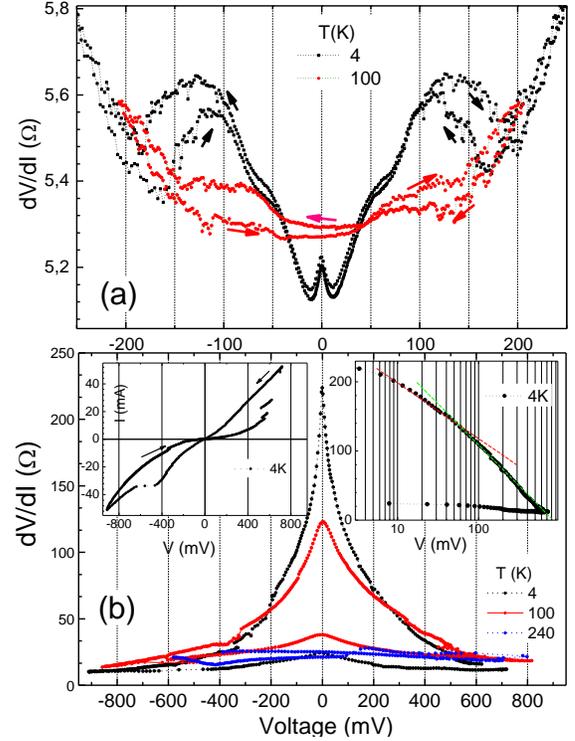

Fig. 5. (a) $dV/dI(V)$ of VSe$_2$ – Ag PC. The inset shows the same $dV/dI(V)$ at a small bias with parabolic fit (dashed red curve). (b) $d^2V/dI^2(V)$ of VSe$_2$ – Ag PC from the inset in the upper panel. The inset shows the same $d^2V/dI^2(V)$ (black dots) after subtracting the linear background shown in the main panel by a dashed (red) line. The solid (red) curve is after subtraction of N-shaped (red dash curve) zero bias anomaly modeled as a derivative of Lorenz distribution (maximum). In this case, a shoulder/kink around 10 mV is more visible. Arrows mark the position of the main peak at around 20 meV and a number of maxima around 10 meV in the EPI spectrum presented in [27].

Fig. 6. (a) $dV/dI(V)$ of VSe$_2$ – Ag PC at helium temperature and at a temperature close to CDW transition. (b) $dV/dI(V)$ of VSe$_2$ – Cu PC with the effect of switching at different temperatures. The left inset shows the $I(V)$ curve at 4 K from the main panel. The right inset shows $dV/dI(V)$ at 4 K in log-scale.

At higher voltages, $dV/dI(V)$ shows broad humps above ±100 mV, which disappears when reaching 100 K (see Fig. 6(a)), that is where resistance inflection connected with CDW transition occurs in $\rho(T)$ [12, 13, 14, 16, 29]. Similar humps connected with the CDW transition were found in our PC investigations of TiSe$_2$ [21]. At even higher voltages, $dV/dI(V)$ demonstrates again metallic upturn. Interestingly, this inflection in $\rho(T)$ transforms into an N-shaped feature with maximum around 80 K as shown in Fig. 2s of the Supplement [15], which grows and shifts to higher temperatures under pressure [16]. So, in general, the behavior of $dV/dI(V)$ resembles $\rho(T)$, which indicates the implementation of the thermal regime at higher voltages. This is expected, considering the very high residual resistivity of about $10^{-4}$ Ω·cm and the short mean free path of electrons in VSe$_2$, which can approach the lattice constant [32]. In fact, we detected the CDW transition in our spectra.

We observed also resistive switching for several VSe$_2$ PCs, which is shown in Fig. 6(b). The voltage range and shape of the switching loops are similar to that observed in other TMDs, such as MoTe$_2$, WTe$_2$, Ta*Me*Te$_4$ (*Me*= Ru, Rh, Ir), TiSe$_2$, TiSeS, Cu$_x$TiSe$_2$, TiTe$_2$, which we investigated in [20, 21, 22]. Such similarity may indicate that the nature of the switching is connected with reversible modification of the crystal structure of VSe$_2$ in the PC core apparently due to displacement of V ions to interlayer under a high electric field. We revealed resistive switching by using Ag or Cu as counter electrodes, which prevents the formation of conductive dendrites in the case of Cu, as is possible in the case of electrochemically active Ag [33]. That is, observed switching is not due to the emergence of dendrites. Unfortunately, we haven't been able to keep PCs stable at temperature rise, to look for the evolution of the switching loop from helium up to room temperature, but, in general, the amplitude of the switching loop is decreasing while temperature increases.

**Conclusion**

We applied Yanson PC spectroscopy method to study the electronic properties of layered VS$_2$ and VSe$_2$ compounds. Both metallic and nonmetallic like phases were observed in the case of VS$_2$. Magnetic phase transition was registered below 25 K for the metallic phase. Because of magnetic phase transition, EPI features were not resolved in PC spectra, while the rare PC spectra, where the magnetic transition was not visible, shows a broad maximum of around 20 mV apparently connected with EPI. We also identified the contribution of phonon modes to EPI in VSe$_2$ around 10 and 20 mV. Additionally, features connected with CDW transition were registered in *dV/dI(V)* of VSe$_2$ above 100 mV. Kondo-like zero-bias maximum is resolved in *dV/dI(V)* of both compounds, evidently due to interlayer vanadium ions. Development of the thermal regime with an increase of voltage is characteristic for PCs of both compounds. Remarkably, reversible resistive switching in PCs on VSe$_2$ was revealed around +/- 400 mV, while in the case of VS$_2$, only instability of PC resistance was seen at high voltage. From this perspective, VSe$_2$ may be attractive for resistive memory applications, nanoscale electronics etc. At the same time, VS$_2$ draw attention as a rare van der Waals TMD with magnetic properties favorable for spintronics. The measured variety of the PC spectra testifies about the strong dependence of electronic properties of both compounds on stoichiometry, induced vacancies, interlayer ions, etc., which allow us to realize and observe different prospective ground states.

**Acknowledgement**

We appreciate S. Gaβ and T. Schreiner for the technical assistance. We would like to acknowledge funding by Alexander von Humboldt (DB, OK, YuN) and Volkswagen Foundation (OK). DB, OK and YuN are also grateful for support by the National Academy of Sciences of Ukraine under project Ф19-5 and are thankful to the IFW Dresden for hospitality. SA acknowledges DFG through Grant No. SA 523/4-1. LH acknowledges financial support from DST-India (DST/WOS-A/PM-83/2021(G)) and National Mission on Interdisciplinary Cyber-Physical Systems (NM-ICPS) of the Department of Science and Technology, Govt. Of India through the I-HUB Quantum

**Literature**

[1] *Two Dimensional Transition Metal Dichalcogenides, Synthesis, Properties, and Applications,* Eds. N. S. Arul & V. D. Nithya, Springer Nature, 2019.

[2] A. H. M. Abdul Wasey, and G. P. Das, *Electronic and magnetic properties of vanadi-um dichalcogenides: A brief overview on theory and experiment,* J. Appl. Phys. 131, 190701 (2022).

[3] Yandong Ma, Ying Dai, Meng Guo, Chengwang Niu, Yingtao Zhu, and Baibiao Huang, *Evidence of the existence of magnetism in pristine $VX_2$ monolayers (X = S, Se) and their strain-induced tunable magnetic properties,* ACS Nano 6, 1695 (2012).

[4] H. Zhang, L. M. Liu, and W. M. Lau, *Dimension-dependent phase transition and magnetic properties of $VS_2$,* J. Mater. Chem. A 1, 10821 (2013).

[5] A. Gauzzi, A. Sellam, G. Rousse, Y. Klein, D. Taverna, P. Giura, M. Calandra, G. Loupias, F. Gozzo, E. Gilioli, F. Bolzoni, G. Allodi, R. De Renzi, G. L. Calestani, P. Roy, *Possible phase separation and weak localization in the absence of a charge-density wave in single-phase* $1T$-$VS_2$**,** Phys. Rev. B **89**, 235125 (2014).

[6] M. Mulazzi, A. Chainani, N. Katayama, R. Eguchi, M. Matsunami, H. Ohashi, Y. Senba, M. Nohara, M. Uchida, H. Takagi, and S. Shin, *Absence of nesting in the charge-density-wave system 1T-$VS_2$ as seen by photoelectron spectroscopy*, Phys. Rev. B **82**, 075130 (2010).

[7] Hicham Moutaabbid, Yann Le Godec, Dario Taverna, Benoıⅽ Baptiste, Yannick Klein, Geneviève Loupias, and Andrea Gauzzi, *High-Pressure Control of Vanadium Self-Intercalation and Enhanced Metallic Properties in* 1T-$V_{1+x}S_2$ *Single Crystals*, Inorg. Chem. **55,** 6481 (2016).

[8] H. Nozaki, Y. Ishizawa, M. Saeki, and M. Nakahira, Electrical properties of $V_5S_8$ single crystals, Physics Letters A 54, 29 (1975).

[9] Qingqing Ji, Cong Li, Jingli Wang, Jingjing Niu, Yue Gong, Zhepeng Zhang, Qiyi Fang, Yu Zhang, Jianping Shi, Lei Liao, Xiaosong Wu, Lin Gu, Zhongfan Liu, and Yanfeng Zhang, *Metallic Vanadium Disulfide Nanosheets as a Platform Material for Multifunctional Electrode Applications*, Nano Lett. **17**, 4908−4916 (2017).

[10] Jingjing Niu *et al.*, *Intercalation of van derWaals layered materials: A route towards engineering of electron correlation*, Chin. Phys. B Vol. 29, No. 9, 097104 (2020).

[11] Dian Li, Xiong Wang, Chi-ming Kan, Daliang He, Zejun Li, Qing Hao, Hongbo Zhao, Changzheng Wu, Chuanhong Jin, and Xiaodong Cui, *Structural Phase Transition of Multilayer* $VSe_2$, ACS Appl. Mater. Interfaces, **12**, 25143 (2020).

[12] Michel Bayard and M. J. Sienko, *Anomalous electrical and magnetic properties of vanadium diselenide*, J. Solid State Chem., **19,** 325 (1976).

[13] C.F. van Bruggen and C. Haas, *Magnetic susceptibility and electrical properties of $VSe_2$ single crystals*, Solid State Commun., **20,** 251 (1976).

[14] Sourabh Barua, M. Ciomaga Hatnean, M. R. Lees & G. Balakrishnan, *Signatures of the Kondo effect in* $VSe_2$, Scientific Reports, **7**, 10964 (2017).

[15] S. Sahoo, U. Dutta, L. Harnagea, A. K. Sood, and S. Karmakar, *Pressure-induced suppression of charge density wave and emergenceof superconductivity in* 1*T*-VSe$_2$, Phys. Rev. B **101**, 014514 (2020).

[16] Dongqi Song, Ying Zhou, Min Zhang, Xinyi He and Xinjian Li, *Structural and Transport Properties* of 1T-VSe$_2$ *Single Crystal Under High Pressures*, Front. Mater. 8:710849 (2021).

[17] Yu. G. Naidyuk, I. K. Yanson, Point-Contact Spectroscopy, *Springer Series in Solid-State Sciences.* 2005, vol 145 (New York: Springer).

[18] Yu. Naidyuk, O. Kvitnitskaya, D. Bashlakov, S. Aswartham, I. Morozov, I. Chernyavskii, G. Fuchs, S.-L. Drechsler, R. Hühne, K. Nielsch, B. Büchner and D. Efremov, *Surface superconductivity in the Weyl semimetal* MoTe$_2$ *detected by point contact spectroscopy*, 2D Mater. **5**, 045014 (2018).

[19] Yu. G. Naidyuk, D. L. Bashlakov, O. E. Kvitnitskaya, S. Aswartham, I. V. Morozov, I. O. Chernyavskii, G. Shipunov, G. Fuchs, S.-L. Drechsler, R. Hühne, K. Nielsch, B. Büchner, D. V. Efremov, *Yanson point-contact spectroscopy of Weyl semimetal* WTe$_2$, 2D Mater. **6**, 045012 (2019).

[20] Yu. G. Naidyuk, D. L. Bashlakov, O.E. Kvitnitskaya, B. R. Piening, G. Shipunov, D. V. Efremov, S. Aswartham and B. Büchner, *Switchable domains in point contacts based on transition metal tellurides*, Phys. Rev. Materials **5**, 084004 (2021).

[21] D. L. Bashlakov, O. E. Kvitnitskaya, S. Aswartham, Y. Shemerliuk, H. Berger, D. V. Efremov, B. Büchner, Yu. G. Naidyuk, *Peculiarities of electron transport and resistive switching in point contacts on* TiSe$_2$, TiSeS *and* Cu$_x$TiSe$_2$, Fiz. Nyzhk. Temp. **5**, 084004 (2021) [Low Temp. Phys. **5**, 084004 (2021)].

[22] O. E. Kvitnitskaya, L. Harnagea, D. V. Efremov, B. Büchner, Yu. G. Naidyuk, *Resistive switching and peculiarities of conductivity of* TiTe$_2$ *point contacts*. DPG Spring Meeting, March 26-31, 2023, Dresden. (to be published).

[23] I. O. Kulik, *Ballistic and nonballistic regimes in point-contact spectroscopy*, Fiz, Nizk. Temp 18, 450 (1992) [Sov. J. Low Temp. Phys. 18, 302 (1992)].

[24] B. I. Verkin, I. K. Yanson, I. O. Kulik, O. I. Shklyarevskii, A. A. Lysykh, Yu. G. Naydyuk, *Singularities in* $d^2V/dI^2$ *dependences of point contacts between ferromagnetic metals*, Solid State Communs., **30**, 215 (1979).

[25] Yu. G. Naidyuk, O. E. Kvitnitskaya, I. K. Yanson, G. Fuchs, K. Nenkov, A. Waelte, G. Behr, D. Souptel, and S.-L. Drechsler, *Point-contact spectroscopy of the antiferromagnetic superconductor* HoNi$_2$B$_2$C *in the normal and superconducting state*, Phys. Rev. B, **76**, p. 014520 (2007).

[26] Yu. G. Naidyuk, O.E. Kvitnitskaya, S. Aswartham, G. Fuchs, K. Nenkov, and S. Wurmehl, *Exploring point-contact spectra of* Ba$_{1−x}$Na$_x$Fe$_2$As$_2$ *in the normal and superconducting states*, Phys. Rev. B **89,** 104512 (2014).

[27] Paulina Majchrzak et al., *Switching of the electron-phonon interaction in 1T-VSe$_2$ assisted by hot carriers,* Phys. Rev. B **103**, L241108 (2021).

[28] Yu. G. Naidyuk, O. I. Shklyarevskii, I. K.Yanson, *Microcontact spectroscopy of dilute magnetic alloys* CuMn *and* CuFe, Fiz. Nizhk. Temp. **8**, 725 (1982) [Sov. J. Low Temp. Phys., **8**, 362 (1982)].

[29] Juhi Pandey and Ajay Soni, *Electron-phonon interactions and two-phonon modes associated with charge density wave in single crystalline* 1T-VSe$_2$, Phys. Rev. B, **101**, 014514 (2020).

[30] J. Zasadzinski, D. M. Burnell, and E. L. Wolf, G. B. Arnold, *Superconducting tunneling study of vanadium,* Phys. Rev. B **25**, 1622 (1982).

[31] G. V. Kamarchuk, A. V. Khotkevich, V. M. Bagatskiĭ, P. Molinie, A. Leblanc, E. Faulques, *Spectroscopy of the electron–phonon interaction in the layered two-dimensional dichalcogenide* 1T–$VSe_2$, Fiz. Nizk. Temp. **27**, 73 (2001) [Low Temp. Phys. **27**, 56 (2001)].

[32] H. Mutka and P. Molinie, *Irradiation-induced defects in layered dichalcogenides: the case of* $VSe_2$, J. Phys. C: Solid State Phys., **15**, 6305-6319 (1982).

[33] Rainer Waser, Regina Dittmann, Georgi Staikov, and Kristof Szot. Redox-based resistive switching memories – Nanoionic mechanisms, prospects, and challenges, Adv. Mater. **21**, 2632–2663 (2009).